\documentclass{elsart}     % for users with LaTeX2e
\input epsf.sty
\usepackage{amsfonts}
\begin{document}
\begin{frontmatter}
\title{Mean field analysis of Williams-Bjerknes type growth}
\author[ANU]{M.~T. Batchelor}
\address[ANU]{Department of Mathematics, School of Mathematical Sciences,\\
Australian National University, Canberra ACT 0200, Australia}
\author[UNSW]{B.~I. Henry}
 and \author[UNSW]{S.~D. Watt}
\address[UNSW]{Department of Applied Mathematics, University of New South
Wales, Sydney NSW 2052, Australia}
\date{\today}
\maketitle
\begin{abstract}
We investigate a class of stochastic growth models involving competition
between two phases in which one of the phases has a competitive
advantage.
The equilibrium populations of the competing phases are
calculated using a mean field analysis.
Regression probabilities for the extinction of the advantaged
phase are calculated in a leading order approximation.
The results of the calculations
 are in good agreement with simulations carried out on a square
lattice with periodic boundaries.

The class of models are variants of the Williams-Bjerknes  model
for the growth of tumours in the basal layer of an epithelium.
In the limit in which only one of the phases is unstable the class
of models reduces to the well known variants of the Eden
model.
\end{abstract}

%\keyword stochastic patterns, mean field, tumour growth 
%\endkeyword

%\PACS{61.43.Hv, 68.90+g}

\end{frontmatter}

\section{Introduction}
Stochastic pattern formation is ubiquitous in science and technology
(see, e.g., \cite{V,BS,BJ,M} and the many references therein).
Some examples include viscous fingering in fluids, 
dendritic crystallization, electro-deposition, 
dielectric breakdown and the growth of bacterial colonies.
It is remarkable that, despite their disparate physical origins,
many such patterns can essentially be described by one of two
generic models.
These models are distinguished by the nature of the stochastic growth
at the interface, which is usually taken to be either:
(i) uniform -- as in the Eden model \cite{E1,E2}, or
(ii) Laplacian -- as in the diffusion-limited aggregation model \cite{WS}.

These generic models and their numerous refinements have been 
intensively studied
and have proved highly successful in describing the growth of patterns
in which one stable phase propagates into a second unstable phase.
For example, in the formation of snowflakes the snow crystal
(stable phase) propagates into water vapour (unstable phase).
In viscous fingering a low viscosity fluid (stable phase)
is pumped under pressure into a background high viscosity fluid
(unstable phase).
These are examples of Laplacian growth processes.
An example of a uniform growth process is the spread of a
bacterial colony (stable phase) via cell-division into a
nutrient-rich environment (unstable phase) \cite{FN1}.

However, many important growth processes have
two or more competing unstable phases.
An example is the Williams-Bjerknes (WB) model \cite{WB}
for the growth of tumours, in which both the cancerous  cells
and the normal cells are unstable (both may be displaced
by cell division). The limiting growth is governed by this competition.
The cancerous cells have the competitive advantage
of dividing faster than
normal cells.
Importantly, early simulations of the model \cite{WB} revealed that the invasive
nature of the tumours can be accounted for solely by an almost
evenly balanced contest between the cancerous cells and the normal cells.
These early simulations also investigated
the regression probability, 
defined as the probability that the advantaged phase,
starting out as a single seed, will become extinct under the growth
process.
Subsequent simulations \cite{DM} and rigorous mathematical studies
\cite{MD,BG}
revealed that the abnormal region, whenever it survives,
can be described as
an asymptotic shape (with a one-dimensional boundary),
the average radius of which grows linearly in time.

In this paper we introduce a class of uniform stochastic growth models
(which includes the WB model) involving competition
between two phases.
The equilibrium populations of the competing phases are
calculated using a mean field analysis similar to that used
recently for a dynamical model of virus spread \cite{CC}.
Leading order approximations for regression probabilities are also
calculated.

The mean field
results and regression results
are in good agreement with simulations carried out on a square
lattice with periodic boundaries.

\section{Competing Growth Models - Mean-Field Equations}
Consider a regular lattice with $N$ sites where
$N_A$ are of type $A$ and $N_B$ are type $B$.
Define an interface site
as one which has one or more nearest 
neighbouring sites of the opposite type.
Let ${\mathcal N}_A$ and ${\mathcal N}_B$ denote the number of type $A$
interface sites and
type 
$B$ interface sites respectively. The total number of nearest neighbour sites
surrounding a lattice site is the lattice co-ordination number denoted
here by $n_c$ (for example, $n_c=4$ on the square lattice).

In the models below the numbers of type $A$ sites and type $B$ sites
change in time as a consequence of interface interactions
which occur
when an interface site
is changed into the opposite type.
Let $p_{i\rightarrow B}$ 
denote the probability for site $i$ to be changed
to a type $B$ site and similarly for $p_{i\rightarrow A}$.
The complete set
of transition probabilities for the site $i$ are then;
\begin{eqnarray}
\label{WBBA}p_{i(B)\rightarrow A}&=&p_{i\rightarrow A}\delta_{i,B},\\
\label{WBAB} p_{i(A)\rightarrow B}&=&p_{i\rightarrow B}\delta_{i,A},\\
\label{WBAA}p_{i(A)\rightarrow A}&=&p_{i\rightarrow A}\delta_{i,A},\\
\label{WBBB}p_{i(B)\rightarrow B}&=&p_{i\rightarrow B}\delta_{i,B}.
\end{eqnarray}
where $i(A)$ and $i(B)$ is used to represent the situation in which
site $i$ is a type $A$ site and a type $B$ site respectively.
The mean field probability that a $B$ site will be changed into an $A$ site
is thus
\begin{equation}
p_{B\rightarrow A}=\sum_i^N p_{i\rightarrow A}\delta_{i,B}.
\end{equation}
Similarly
\begin{equation}
p_{A\rightarrow B}=\sum_i^N p_{i\rightarrow B}\delta_{i,A}
\end{equation}
is the
mean field probability that an $A$ site will be changed into a $B$ site.
Using these
probabilities we can write down
general mean field population equations for this class
of models which relates the numbers of species before
(time $t$) 
and after (time $t+1$) an interface interaction:
\begin{eqnarray}
N_A(t+1)&=&p_{B\rightarrow A}(t)-p_{A\rightarrow B}(t)+N_A(t),\\
N_B(t+1)&=&p_{A\rightarrow B}(t)-p_{B\rightarrow A}(t)+N_B(t).
\end{eqnarray}
The mean field steady state defined by
$N_A(t+1)=N_A(t)=N_A^{*}$
and $N_B(t+1)=N_B(t)=N_B^{*}$
thus yields the (equivalent) equilibrium conditions
\begin{eqnarray}
\label{eqm1} p_{B\rightarrow A}^{*}&=&
p_{A\rightarrow B}^{*},\\
\sum_i^N p_{i\rightarrow A}^{*} \delta_{i,B}&=&
\label{eqm2} \sum_i^N p_{i\rightarrow B}^{*} \delta_{i,A}.
\end{eqnarray}
We shall make use of both forms of the mean-field equilibrium condition
in the following.
Note that
the steady state mean field probabilities
 $p_{B\rightarrow A}^{*}$ and $p_{A\rightarrow B}^{*}$ are in general
 functions of $N_A^{*},{\mathcal N}_A^{*},
N_B^{*},{\mathcal N}_B^{*}$.

\section{ Williams-Bjerknes Model}
In the WB model,
the spread of cancers is modelled by
the growth of two types of cells,
cancerous cells and normal cells, which compete via cell division
in a basal layer.
The cancerous cells are taken to divide $\kappa$ times faster than normal cells.
This factor $\kappa$ is called the carcinogenic advantage.
It is supposed that
when a cell in the 
basal layer divides the daughter cell
displaces a neighbouring cell up out of the basal layer
\cite{FN2}.
The implementation of the model is straightforward.
Start with a regular lattice of sites (basal layer)
 labelled type $A$ (cancerous) or type $B$
(normal).
Select a site at random
 with a bias (due to the carcinogenic advantage)
for selecting type $A$.
Choose one of the neighbouring sites of the selected site
 at random and convert it
to the same type as the selected site.
Typically there are many growth steps
that do not change the configuration of the basal layer.
%
% REVIEW AFTER DISCUSSION
%Such non-configurational changes cannot be ignored
%since they play a role
%in both the rate of approach to equilibrium
% and in the nature of the equilibrium
%state.

\begin{figure}
\centerline{
\epsfxsize=5.5in
\epsfbox{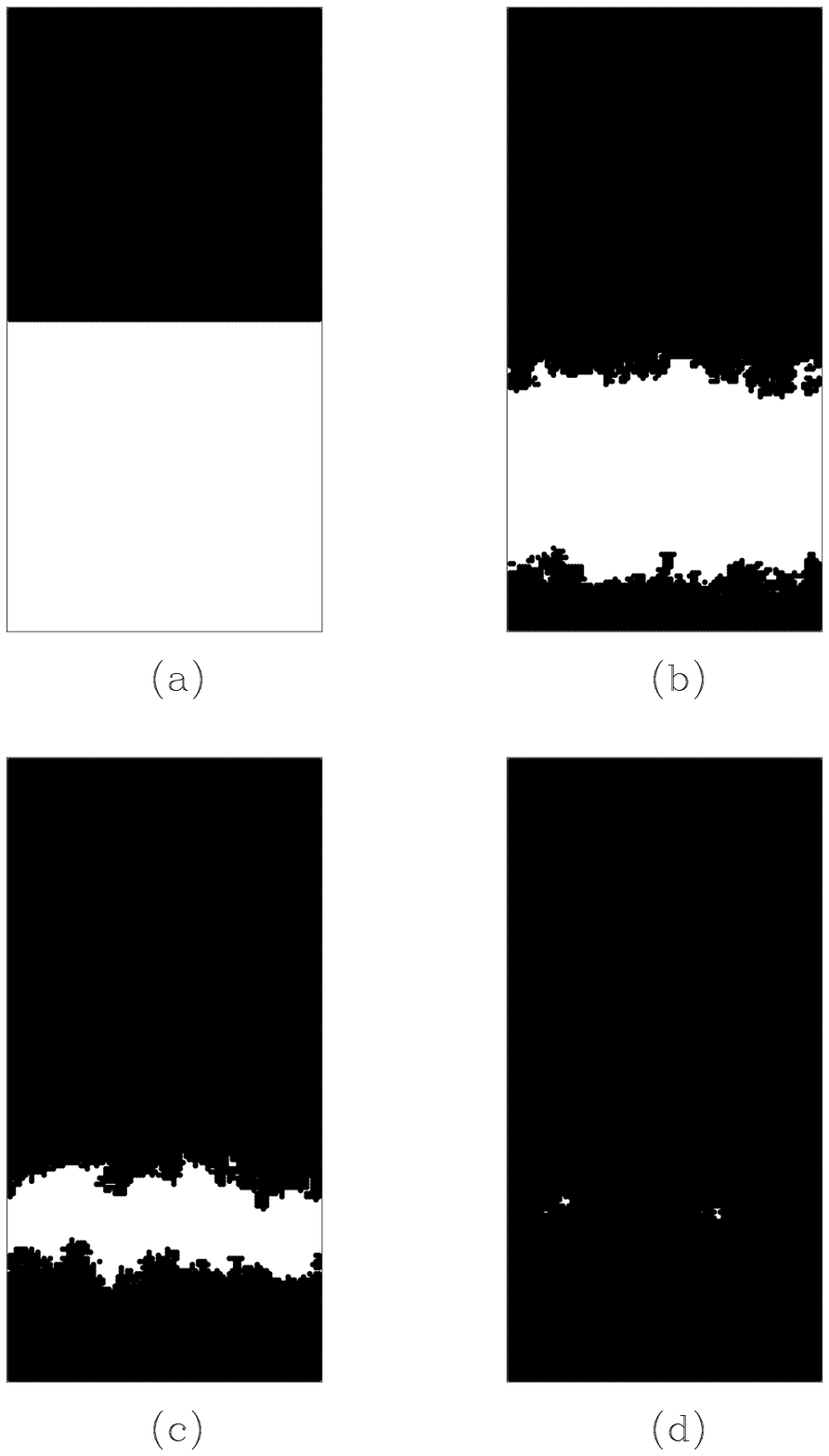}}
\caption{Configurations of normal cells (white) and abnormal cells
(black) in the WB model with carcinogenic advantage $\kappa=2$;
(a) initially, (b) after $8\times 10^5$ steps, (c) after $16\times 10^5$ 
steps, (d) after $24\times 10^5$ steps.}
\label{fig1}
\end{figure}

We have simulated this model within a rectangular region
covering $[0,100]\times [-100,100]$ 
square lattice sites using periodic
boundaries in both directions. Initially the top
half of the rectangular region is comprised of abnormal cells and
 the bottom half is comprised of the normal cells.
The configurations of normal and abnormal cells
for the case $\kappa=2$ are shown
at four different time snapshots in Fig.\ \ref{fig1}.
Two points of note are: i) the interface between normal
and abnormal cells is highly irregular;
ii) the ratio abnormal cells to normal
cells increases in time until (on average) the advantaged
species completely dominates.

\subsection{Mean Field Equilibrium}
To determine the mean field equilibrium state for this model we first define
the model transition probabilities.
The probability that a site $i$ will be changed ($B\rightarrow A$)
or reconfigured ($A\rightarrow A$) into type $A$ is
\begin{equation}
p_{i\rightarrow A}=
\left({\kappa N_A\over\kappa N_A+N_B}\right)
\left({n_i(A)\over N_A}\right)
\left({1\over n_c}\right), \label{WBiA}
\end{equation}
where $n_i(A)$ is the number of neighbouring $A$ sites adjoining
site $i$.
The first factor is the carcinogenic advantage for a division
of an $A$ cell. The second factor
is the probability that the $A$ cell chosen for division is
one of the neighbours of site $i$.
The final factor is the probability that if a neighbouring $A$ cell
for site $i$ is chosen for division it will displace the
cell at site $i$ out of the basal layer leaving an $A$ cell at that location.
Similarly the probability that a site $i$ will be changed or reconfigured
into type $B$ is
\begin{equation}
p_{i\rightarrow B}=\left({N_B\over\kappa N_A+N_B}\right)
\left({n_i(B)\over N_B}\right)
\left({1\over n_c}\right). \label{WBiB}
\end{equation}
The normalization of the probabilities,
\begin{equation}
P=\sum_i^N\left(p_{i\rightarrow A}+p_{i\rightarrow B}\right)=1,
\end{equation}
follows immediately from the identities
\begin{eqnarray}
\label{identityA} \sum_i^N \frac{n_i(A)}{n_c}&=&N_A,\\
\sum_i^N \frac{n_i(B)}{n_c}&=&N_B. \label{identityB}
\end{eqnarray}

\begin{figure}
\centerline{
\epsfxsize=3.5in
\epsfbox{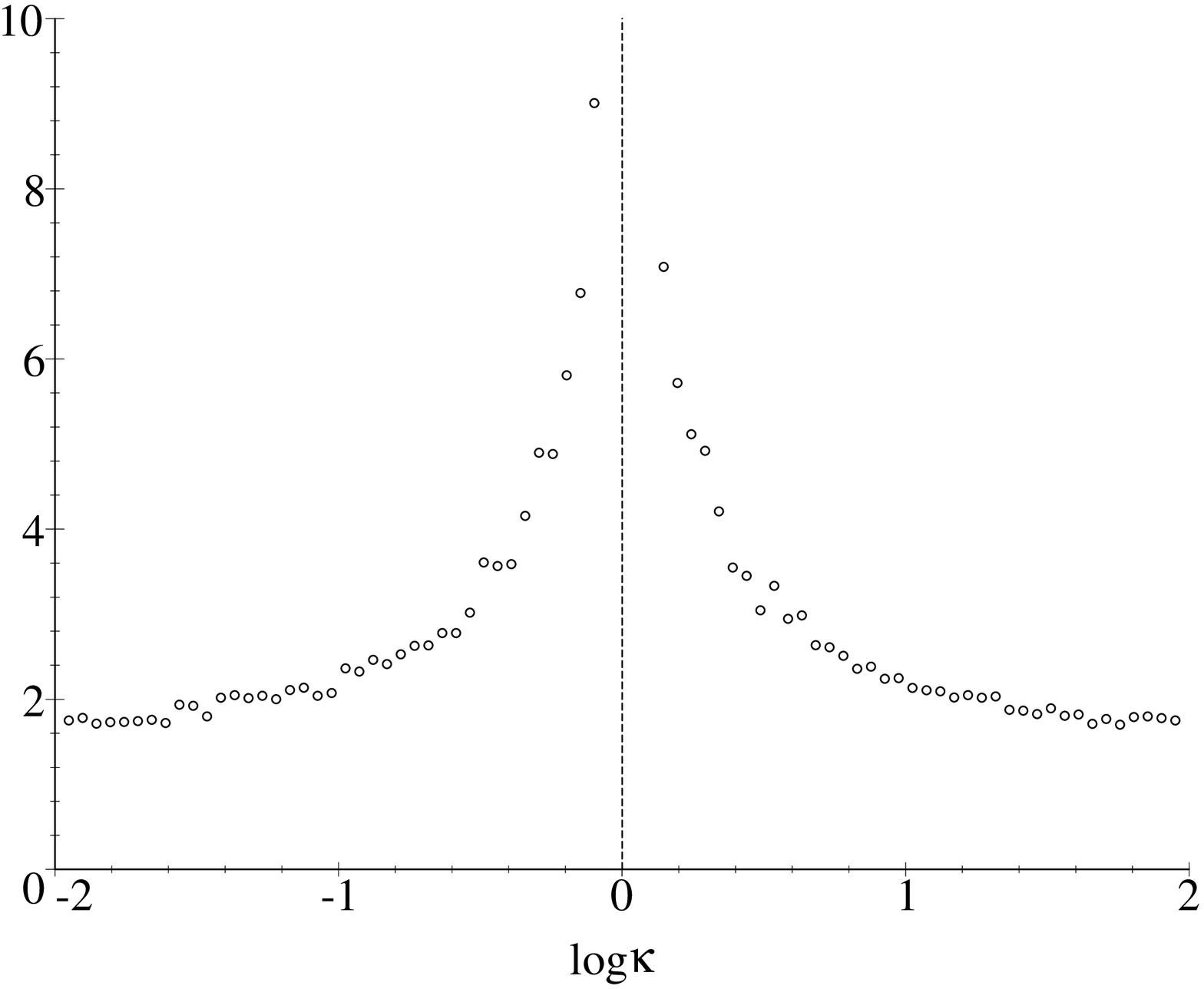}}
\caption{Time taken for the extinction of normal cells
($\kappa>1$) and the extinction of abnormal cells
($\kappa<1$) as a function of $\log \kappa$
in the WB model.}
\label{fig2}
\end{figure}

Substituting the expressions for the transition
probabilities (\ref{WBiA}), (\ref{WBiB})
 into the mean field equilibrium condition (\ref{eqm2})
yields
\begin{equation}
\kappa\sum_i^{N}
n_i(A)\delta_{i,B}=
\sum_i^N
n_i(B)\delta_{i,A}. \label{WBE1}
\end{equation}
This result can be simplified by the identity
\begin{equation}
\label{identity} \sum_i^N n_i(A)\delta_{i,B}=\sum_i^N n_i(B)\delta_{i,A};
\end{equation}
which follows from
the Eqs (\ref{identityA}), (\ref{identityB}), together with the 
neighbour summation rule $n_i(A)+n_i(B)=n_c$.
Using the identity (\ref{identity})
the mean field equilibrium result (\ref{WBE1})
simplifies to
\begin{equation}
\label{WBeqm} \kappa=
1.
\end{equation}
From our stochastic simulations
described above we have have obtained a plot
in  Fig.\ \ref{fig2} of the time
taken for all $B$ cells to die out ($\kappa>1$) and the time for all $A$ cells
to die out ($\kappa<1$) as a function of $\log\kappa$.
The extinction time
rapidly increases with decreasing $|\kappa-1|$ and there is a clear
possibility for a steady state with both species at $\kappa=1$.

A different form of the equilibrium condition
can be derived by writing
the mean field
probability that any $B$ type cell is changed into an $A$ type cell
as 
\begin{equation}
p_{B\rightarrow A}=
\left({\kappa N_A\over\kappa N_A+N_B}\right)
\left(\frac{\mathcal N_A}{N_A}\right)
\alpha_{B\rightarrow A}. \label{WBBAnew}
\end{equation}
The first factor on the right is
the probability that an $A$ cell
 divides, the second factor is the
probability that the dividing $A$ cell is an interface cell
and the third factor is
the probability that $A$ interface cells
displace $B$ interface cells (there is a finite
probability that $A$ cells might divide without any resultant
change in the configuration).
Similarly the mean field probability that any $A$ type cell is changed into a $B$ 
type cell can be written as
\begin{equation}
p_{A\rightarrow B}=
\left({N_B\over\kappa N_A+N_B}\right)
\left(\frac{{\mathcal N}_B}{N_B}\right)
\alpha_{A\rightarrow B}. \label{WBABnew}
\end{equation}
Substituting (\ref{WBBAnew}) and (\ref{WBABnew}) into
the mean field equilibrium condition (\ref{eqm1})
now yields
\begin{equation}
\kappa{\mathcal N}_A\alpha_{B\rightarrow A}
=
{\mathcal N}_B\alpha_{A\rightarrow B}. \label{WBeqm1}
\end{equation}
Comparing (\ref{WBeqm1}) with the previous equilibrium condition (\ref{WBeqm})
 we obtain the additional
equilibrium result
\begin{equation}
\frac{\alpha_{A\rightarrow B}}{\alpha_{B\rightarrow A}}=
\frac{{\mathcal N}_A}{{\mathcal N}_B}. \label{WBeqm2}
\end{equation}

Note that in general we can write
\begin{equation}
\label{E1} \sum_i^N n_i(B)\delta_{i,A}=\alpha_{B}{\mathcal N_B},
\end{equation}
where the factor $\alpha_{B}$ is included because some
of the $B$ interface cells are counted more than once in the sum,
and similarly
\begin{equation}
\label{E2} \sum_i^N n_i(A)\delta_{i,B}=\alpha_{A}{\mathcal N_A}.
\end{equation}
Combining (\ref{E1}) and (\ref{E2}) with the identity (\ref{identity}) thus
yields
\begin{equation}
\label{E3} \alpha_{A}{\mathcal N}_A=\alpha_{B}{\mathcal N}_B.
\end{equation}
Comparing the general result (\ref{E3})
with the equilibrium condition (\ref{WBeqm2})
 we now have
\begin{equation}
\frac{\alpha_{A}}{\alpha_{B}}=\frac{\alpha_{B\rightarrow A}}
{\alpha_{A\rightarrow B}}.
\end{equation}

\subsection{Regression}
Here we consider the case in which
a single cancerous site ($N_A=1$) is
 surrounded by normal sites ($N_B=N-1$).
The transition probabilities are thus
\begin{eqnarray}
p_{i(B)\rightarrow A}&=&
\left({\kappa\over\kappa+N-1}\right)n_i(A)\frac{1}{n_c}\delta_{i,B},\\
p_{i(A)\rightarrow B}&=&
\left({N-1\over\kappa+N-1}\right)\left({n_i(B)\over N-1}\right)
\frac{1}{n_c}\delta_{i,A},\\
p_{i(A)\rightarrow A}&=&
\left({\kappa\over\kappa+N-1}\right)n_i(A)\frac{1}{n_c}\delta_{i,A},\\
p_{i(B)\rightarrow B}&=&
\left({N-1\over\kappa+N-1}\right)\left({n_i(B)\over N-1}\right)\frac{1}{n_c}
\delta_{i,B}.
\end{eqnarray}
Summing over the lattice sites we have
\begin{eqnarray}
p_{B\rightarrow A}&=&
\left({\kappa\over\kappa+N-1}\right),\\
p_{A\rightarrow B}&=&
\left({1\over\kappa+N-1}\right),\\
p_{A\rightarrow A}&=&0,\\
p_{B\rightarrow B}&=&
\left({N-2\over\kappa+N-1}\right).
\end{eqnarray}

The probability $p_{A\rightarrow B}$ is the probability for regression in one
step. 
The probability for regression without any growth within $m$ steps
is given by
\begin{eqnarray}
p_{A\rightarrow B}(m)&=&\sum_{k=1}^m (p_{B\rightarrow B})^{k-1}
 \label{regressm} p_{A\rightarrow B},\\
&=&\frac{1-\left(\frac{N-2}{\kappa+N-1}\right)^m}{\kappa+1},
\end{eqnarray}
and the probability for regression without growth
after infinitely many steps is
\begin{eqnarray}
p_{A\rightarrow B}(\infty)&=&\sum_{k=1}^\infty (p_{B\rightarrow B})^{k-1}
 p_{A\rightarrow B},\\
&=&\frac{1}{\kappa+1}. \label{WBregress}
\end{eqnarray}

We anticipate that the probability for regression without
growth will be the dominant term
in the overall probability for regression, which includes
the probability for regression after growth,
so that (\ref{WBregress}) may be taken as a leading order approximation.
To investigate this further consider regression within three steps.
The possible scenarios and associated probabilities are:
\begin{eqnarray}
A\rightarrow B\, ; B\rightarrow B\, ; B\rightarrow B&:
\qquad &
p_{A\rightarrow B};\\
B\rightarrow B\, ; A\rightarrow B\, ; B\rightarrow B&:
\qquad &
p_{B\rightarrow B}p_{A\rightarrow B};\\
B\rightarrow B\, ; B\rightarrow B\, ; A\rightarrow B&:
\qquad &
(p_{B\rightarrow B})^2p_{A\rightarrow B};\\
B\rightarrow A\, ; A\rightarrow B\, ; A\rightarrow B&:
\qquad &
p_{B\rightarrow A}\hat p_{A\rightarrow B}p_{A\rightarrow B};
\end{eqnarray}
The sum of the probabilities in the first three scenarios above
 is the probability
for regression without growth within three steps
(Eq (\ref{regressm}) with $m=3$)
and
\begin{equation}
\hat p_{A\rightarrow B}=\left(\frac{3}{2}\right)\left(\frac{N-2}{2\kappa+N-2}\right)\left(\frac{1}{N-2}\right)
\end{equation}
is the probability $A\rightarrow B$ in a cluster with two adjacent type
$A$ cells.
It immediately follows that the ratio the probability
 for regression after growth within three steps to the
 probability for regression without
growth within three steps scales as $1/N^3$.

\begin{figure}
\centerline{
\epsfxsize=3.5in
\epsfbox{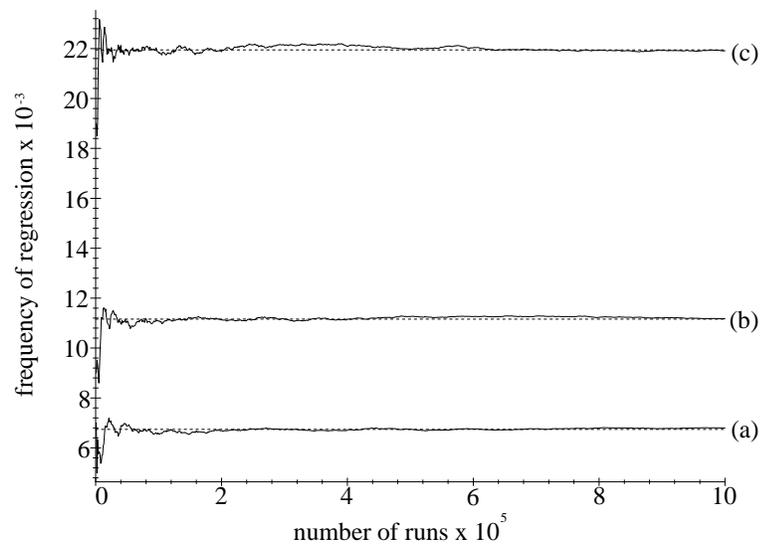}}
\caption{
The frequency
of regression within (a) three (b) five and (c) ten steps
versus the total number of runs used in the frequency
estimate for the WB model.
The lower horizontal line is the probability of regression after three
steps and the two upper horizontal lines are the probabilities for regression
without growth after (b) five and (c) ten steps.
}
\label{fig3}
\end{figure}

We have studied regression in simulations 
with $\kappa=2$ over $[-10,10]\times [-10,10]$ sites on a square lattice.
In Fig.\ \ref{fig3} we have plotted the frequency 
of regression within three, five and ten steps
versus the total number of runs used in the frequency
estimate, over a range of the total number of runs up to $10^6$.
The upper two horizontal lines on the plot are the probabilites for regression
without growth after five and ten steps and the lower horizontal line
is the full probability for regression after three steps.
The comparison between the Monte Carlo results and the
algebraic regression results
provides evidence that regression without growth dominates the
regression process.

\subsection{Limit $\kappa\rightarrow\infty$}
In the 
limit $\kappa\rightarrow\infty$ the $B$ cells
never divide and the transition
probabilities become:
\begin{eqnarray}
p_{i(A)\rightarrow B}&=&0,\\
p_{i(B)\rightarrow B}&=&0,\\
p_{i(A)\rightarrow A}&=&\left({n_i(A)\over N_A}\right)\frac{1}{n_c}\delta_{i,A},\\
p_{i(B)\rightarrow A}&=&\left({n_i(A)\over N_A}\right)\frac{1}{n_c}\delta_{i,B}.
\end{eqnarray}

In this case there are only two possible events $A\rightarrow A$
and $B\rightarrow A$.
The event
$A\rightarrow A$
affects the time scale for configurational changes but
does not affect the configurations themselves.
Hence if we do not concern ourselves
with the time scales for change we can
set $p_{i(A)\rightarrow A}=0$ and renormalize 
\begin{equation}
\label{Emodel} p_{i(B)\rightarrow A}=\frac{n_i(A)\delta_{i,A}}{\sum_i^{\mathcal N_B}n_i(A)}
\end{equation}
where the sum is over all type $B$ interface sites.
In this limit the model
is equivalent to the original model introduced by Eden
\cite{E1,E2}, which is also referred to as the
Eden B model \cite{JB}.

\section{Interface Model}
In this model only interface sites are considered for division.
Select an interface site at random with a bias for selecting type $A$ 
and convert one of the neighbouring sites of the selected site
to a site of the same type.
Note that this model still allows non-configurational changes
where the number of type $A$ interface sites and the number of
type $B$ interface sites may both be conserved
after a growth event.
Let $\bar n_i(A)$ denote the number of type $A$ interface sites
adjoining site $i$ and let $\bar n_i(B)$ denote the number of type
 $B$ interface sites
adjoining site $i$.
The transition probabilities for site $i$ to change to an $A$
site or a $B$ site are
\begin{eqnarray}
\label{IiA} p_{i\rightarrow A}&=&
\frac{\kappa{\mathcal N_A}}{\kappa{\mathcal N_A}+{\mathcal N_B}}
\frac{\bar n_i(A)}{\mathcal N_A}\frac{1}{n_c},\\
\label{IiB} p_{i\rightarrow B}&=&
\frac{{\mathcal N_B}}{\kappa{\mathcal N_A}+{\mathcal N_B}}
\frac{\bar n_i(B)}{\mathcal N_B}\frac{1}{n_c}.
\end{eqnarray}
\subsection{Mean Field Equilibrium}
Substituting (\ref{IiA},\ref{IiB}) into the equilibrium condition (\ref{eqm2})
yields
\begin{equation}
{\kappa}\sum_i^N \bar n_i(A)\delta_{i,B}=\sum_i^N \bar n_i(B)\delta_{i,A}.
\end{equation}
We can simplify this expression by noting that
all type $A$ sites adjoining
 a type $B$ site are interface sites and vice-versa hence we have the identities
\begin{eqnarray}
\sum_i^N\bar n_i(A)\delta_{i,B}&=&\sum_i^N n_i(A)\delta_{i,B},\\
\sum_i^N\bar n_i(B)\delta_{i,A}&=&\sum_i^N n_i(B)\delta_{i,A}.
\end{eqnarray}
Substituting the above identities together with the identity
(\ref{identity})
we obtain the equilibrium condition
\begin{equation}
\kappa=1.
\end{equation}
For other values of $\kappa$ it is expected that
the system will evolve until all sites are the same type;
type $A$ for $\kappa>1$ and
type $B$ for $\kappa<1$.

\begin{figure}
\centerline{
\epsfxsize=5.5in
\epsfbox{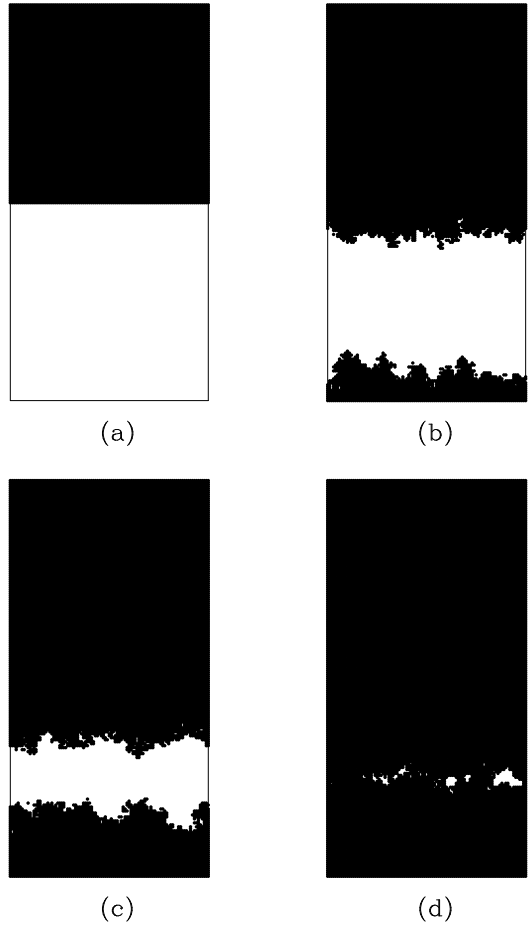}}
\caption{Configurations of normal cells (white) and abnormal cells
(black)
in the Interface Model model with carcinogenic advantage $\kappa=2$;
(a) initially, (b) after $25\times 10^3$ steps, (c) after $50\times 10^3$ 
steps, (d) after $75\times 10^3$ steps.}
\label{fig4}
\end{figure}

The mean field equilibrium results have been found to provide a good
description of the long time populations in stochastic simulations of
this model carried out on
$[0,100]\times[-100,100]$ square lattice sites with
the top half of the region initially comprised of abnormal
cells and the bottom half initially comprised of normal cells.
In Fig.\ \ref{fig4} we show four different time snapshots for the 
case $\kappa=2$.

\begin{figure}
\centerline{
\epsfxsize=3.5in
\epsfbox{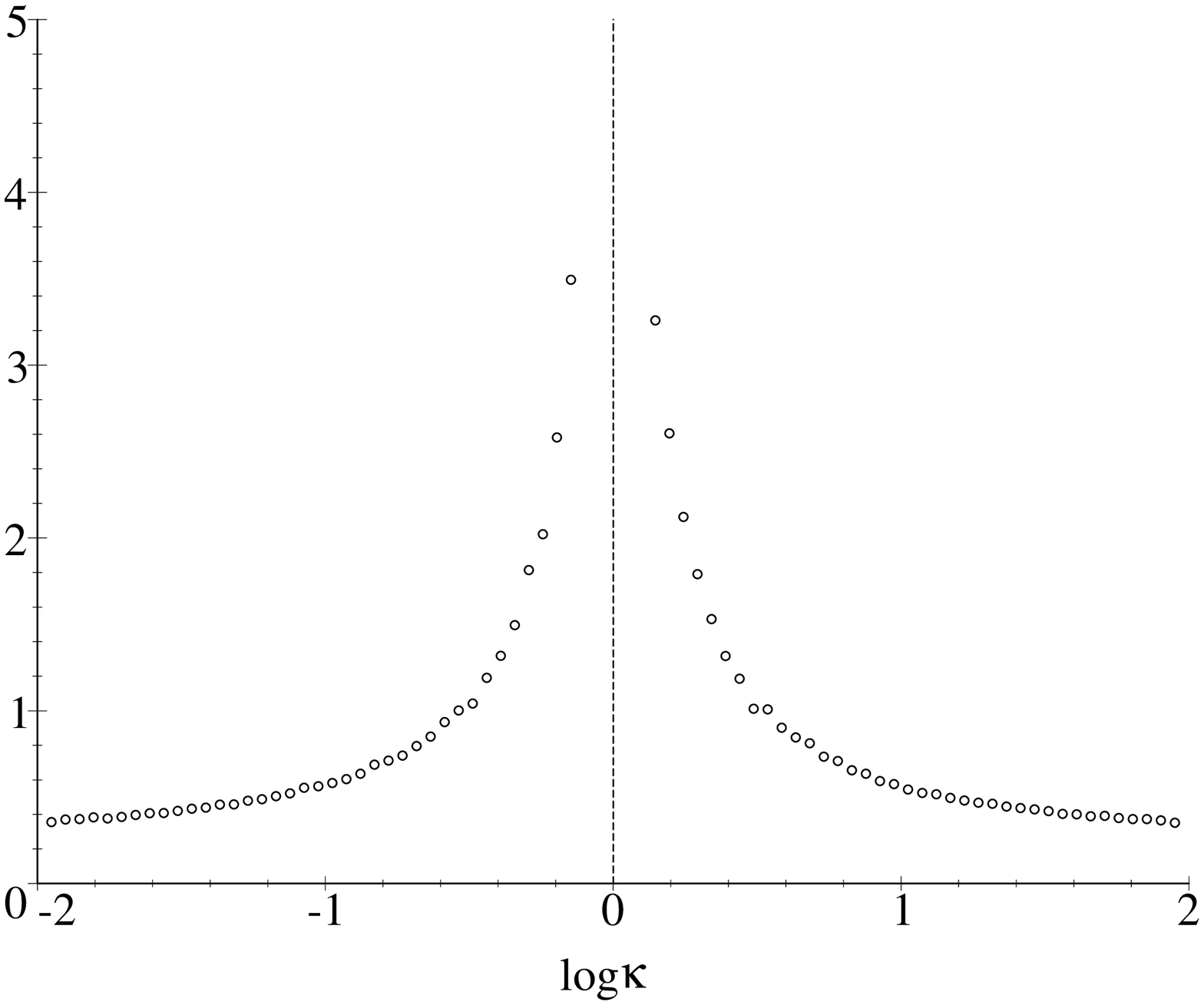}}
\caption{Time taken for the extinction of normal cells
($\kappa>1$) and the extinction of abnormal cells
($\kappa<1$) as a function of $\log \kappa$
in the Interface Model.}
\label{fig5}
\end{figure}

As in the WB model the interface is highly irregular and the advantaged
species completely dominates.  However, the time taken for the extinction
of the disadvantaged species is considerably less than for the WB model.
The extinction times for the interface model over a range of $\kappa$
are shown in Fig.\ \ref{fig5}.
As in the original WB model the
extinction time
rapidly increases with decreasing $|\kappa-1|$ and there is
again a clear
possibility for a steady state with both species at $\kappa=1$.
Comparing Fig.\ \ref{fig2} and Fig.\ \ref{fig5} we see that for
$\approx 10^4$ cells the extinction time is an order of magnitude
less in the interface model than in the WB model.

\subsection{Regression}
In the case where initially we have a single type $A$ site surrounded by
type $B$ sites the transition probabilities are
\begin{eqnarray}
p_{B\rightarrow A}&=&\frac{\kappa}{\kappa+4},\\
p_{A\rightarrow B}&=&\frac{1}{\kappa+4},\\
p_{A\rightarrow A}&=&0,\\
p_{B\rightarrow B}&=&\frac{3}{\kappa+4}.
\end{eqnarray}
The probability for regression without growth after $m$ steps is
\begin{eqnarray}
p_{A\rightarrow B}(m)&=&\sum_{k=1}^m\left(\frac{3}{\kappa+4}\right)^{k-1}\frac{1}{(\kappa+4)},\\
&=&\frac{1-\left(\frac{3}{\kappa+4}\right)^m}{\kappa+1}.
\end{eqnarray}
The probability for regression without growth after infinitely many
 steps is again
\begin{equation}
p_{A\rightarrow B}(\infty)=\frac{1}{\kappa+1}.
\end{equation}
\subsection{Limit $\kappa\rightarrow\infty$}
The $\kappa\rightarrow\infty$ limit of this interface model is equivalent
to the $\kappa\rightarrow\infty$ limit of the WB model
and is thus equivalent to the Eden B model.
\section{Selective Interface Model}
In the selective interface model an interface site is selected at random
with a bias for selecting a type $A$ site.
Then one of the {\sl interface} neighbouring
sites of the selected site is chosen at random
and  converted to the same type
as the initially selected site.
The transition probabilities are
\begin{eqnarray}
p_{i\rightarrow A}&=&\frac{\kappa{\mathcal N_A}}{\kappa{\mathcal N_A}+{\mathcal N_B}}\frac{1}
{\mathcal N_A}\sum_{i'}^{n_c}
\frac{\delta_{i',A}}{n_{i'}(B)},\\
p_{i\rightarrow B}&=&
\frac{{\mathcal N_B}}{\kappa{\mathcal N_A}+{\mathcal N_B}}
\frac{1}{\kappa+1}\frac{1}
{\mathcal N_B}\sum_{i'}^{n_c}
\frac{\delta_{i',B}}{n_{i'}(A)},
\end{eqnarray}
where the sum $\sum_{i'}^{n_c}$ denotes a sum over the 
neighbours of site $i$.

\subsection{Mean Field Equilibrium}
The mean
field equilibrium condition (\ref{eqm2}) for this model
is
\begin{equation}
{\kappa}\sum_i\sum_{i'}^{n_c}\frac{\delta_{i',A}}{n_{i'}(B)}
\delta_{i,B}
=
\sum_i\sum_{i'}^{n_c}\frac{\delta_{i',B}}{n_{i'}(A)}
\delta_{i,A}.
\end{equation}
This expression can be simplified. First we reverse the order
of the
sum over interface sites, $\sum_i$, and
the sum over neighbours, $\sum_i'$,  to obtain
\begin{equation}
{\kappa}\sum_i\sum_{i'}^{n_c}
\delta_{i',B}
\frac{\delta_{i,A}}{n_{i}(B)}
=
\sum_i\sum_{i'}^{n_c}
\delta_{i',A}
\frac{\delta_{i,B}}{n_{i}(A)}.
\end{equation}
It is now a trivial matter to perform the sum over neighbours;
\begin{eqnarray}
\sum_{i'}^{n_c}\delta_{i',B}&=&n_i(B),\\
\sum_{i'}^{n_c}\delta_{i',A}&=&n_i(A),
\end{eqnarray}
leading to the  equilibrium condition
\begin{equation}
{\kappa}{\mathcal N_A}
=
{\mathcal N_B}.
\end{equation}
In general we might expect that the number of type $A$ interface sites
and the number of type $B$ interface sites are approximately equal.
The mean field steady state equilibrium
that is consistent with this expectation is;
$\kappa=1$ and ${\mathcal N_B}={\mathcal N_A}$,
or 
${\mathcal N_B}={\mathcal N_A}=0$.
The result ${\mathcal N_B}={\mathcal N_A}=0$ is
consistent with all type $A$ sites for $\kappa>1$ and all
type $B$ sites for $\kappa<1$.

\subsection{Regression}
In the case where a single type $A$ site is surrounded by $n_c$
type $B$ sites the transition probabilities reduce to
\begin{eqnarray}
p_{B\rightarrow A}&=&\frac{\kappa}{\kappa+4},\\
p_{A\rightarrow B}&=&\frac{4}{\kappa+4},
\end{eqnarray}
so that the probability for regression without growth in this
model is $\frac{4}{\kappa+4}$.

\subsection{Limit $\kappa\rightarrow\infty$}
In the 
limit $\kappa\rightarrow\infty$ the 
transition probability is
\begin{equation}
p_{i\rightarrow A}=\frac{1}
{\mathcal N}_A\sum_{i'}^{n_c}
\frac{\delta_{i',A}}{n_{i'}(B)},
\end{equation}
which is equivalent to the Eden C model \cite{JB}.

\section{Cross-Feeding Model}
Consider a colony of $\mathcal N_A$ type $A$ interface cells and $\mathcal N_B$ type
$B$ interface cells. Select a cell at random with a bias for choosing
type $B$. Convert the chosen cell to the opposite type.
The transition probabilities are
\begin{eqnarray}
p_{i(B)\rightarrow A}&=&
\frac{\kappa{\mathcal N_A}}{\kappa{\mathcal N_A}+{\mathcal N_B}}
\frac{1}{\mathcal N_B}\delta_{i,B},\\
p_{i(A)\rightarrow B}&=&
\frac{{\mathcal N_B}}{\kappa{\mathcal N_A}+{\mathcal N_B}}
\frac{1}{\mathcal N_A}\delta_{i,A}.
\end{eqnarray}
Hence
\begin{eqnarray}
p_{B\rightarrow A}&=&\frac{\kappa{\mathcal N_A}}{\kappa{\mathcal N_A}+{\mathcal N_B}},\\
p_{A\rightarrow B}&=&\frac{{\mathcal N_B}}{\kappa{\mathcal N_A}+{\mathcal N_B}},
\end{eqnarray}
so
 that
the probability for regression without growth is
$\frac{{\mathcal N_B}}{\kappa{\mathcal N_A}+{\mathcal N_B}}$ and the
equilibrium condition is
$\kappa{\mathcal N_A}={\mathcal N_B}$.

In the $\kappa\rightarrow\infty$ limit the probability for a
particular $B$ cell to become an $A$ cell is $1/{\mathcal N}_B$
which is equivalent to the Eden A model \cite{JB}.

\section{Cross-Feeding Model with Neighbour Bias}
Consider a colony of $\mathcal N_A$ type $A$ interface cells and $\mathcal N_B$ type
$B$ interface cells. Select either the set of $A$ cells or the set of $B$ cells
with a bias for choosing the set of $B$ cells.
Pick a cell at random from the selected set with a weighting according
to the number of neighbouring cells of opposite type.
Convert the chosen cell to the opposite type.
The transition probabilities are
\begin{eqnarray}
p_{i(B)\rightarrow A}&=&
\frac{\kappa{\mathcal N_A}}{\kappa{\mathcal N_A}+{\mathcal N_B}}
\frac{n_i(A)}{\mathcal N_B}\delta_{i,\bar B},\\
p_{i(A)\rightarrow B}&=&
\frac{{\mathcal N_B}}{\kappa{\mathcal N_A}+{\mathcal N_B}}
\frac{n_i(B)}{\mathcal N_A}\delta_{i,\bar A},
\end{eqnarray}
where the $\bar B$ denotes an interface $B$ site and the $\bar A$
denotes an interface $A$ site.
In this model the transition probabilities are not automatically
normalized. However, it is a simple matter to calculate the normalization
\begin{eqnarray}
S&=&\frac{1}{\kappa{\mathcal N_A}+{\mathcal N_B}}\left(\frac{\kappa{\mathcal N_A}}{\mathcal N_B}+\frac{{\mathcal N_B}}{\mathcal N_A}\right)
\sum_i^N n_i(A)\delta_{i,\bar B},\\
&=&\frac{1}{\kappa{\mathcal N_A}+{\mathcal N_B}}\left(\frac{\kappa{\mathcal N_A}}{\mathcal N_B}+\frac{{\mathcal N_B}}{\mathcal N_A}\right)
\sum_i^N n_i(B)\delta_{i,\bar A},
\end{eqnarray}
where we have used the identity
\begin{equation}
\label{Id3} \sum_i^N n_i(A)\delta_{i,\bar B}=\sum_i^N n_i(B)\delta_{i,\bar A},
\end{equation}
which
follows immediately from the identity (\ref{identity}).

Summing over all sites we obtain the mean field
 probability that a $B$ cell becomes an $A$ cell and vice-versa:
\begin{eqnarray}
p_{B\rightarrow A}&=&
\frac{\kappa \frac{\mathcal N_A}{\mathcal N_B}}{\kappa\frac{\mathcal N_A}{\mathcal N_B}+
\frac{\mathcal N_B}{\mathcal N_A}},\\
p_{A\rightarrow B}&=&
\frac{\frac{\mathcal N_B}{\mathcal N_A}}{\kappa\frac{\mathcal N_A}{\mathcal N_B}+
\frac{\mathcal N_B}{\mathcal N_A}},\\
\end{eqnarray}
The mean field equilibrium condition in this model is thus
 $\kappa{\mathcal N_A}^2={\mathcal N_B}^2$.
Again if we use the approximation ${\mathcal N_A}\approx {\mathcal N_B}$
then the equilibrium condition is satisfied by;
either $\kappa=1$ and ${\mathcal N_A}\approx {\mathcal N_B}$, 
or ${\mathcal N_A}={\mathcal N_B}=0$.
The latter case occurring when one or other species completely dominates.

In this model if we start with a single type $A$ cell surrounded
by $n_c$ type $B$ cells it is a simple matter
to show that the probability for regression without growth
is $\frac{4}{\kappa+4}$.

The $\kappa\rightarrow\infty$ limit of this model reduces to the original Eden
model \cite{E1,E2}, as in (\ref{Emodel}) above.
\section{Discussion}
In this paper we have introduced a class of stochastic growth models
involving competition between two phases in which one of the phases
has a competitive advantage $\kappa$. This class of models
includes the WB model for tumour growth.
In the limit $\kappa\rightarrow\infty$ where only one of the phases
is unstable the class of models reduces to well known variants of the Eden
model.
We have derived mean field equilibrium results 
and regression probabilities (expressing the probability that
the advantaged phase dies out) for the class of competitive growth models.
The results are found to be in good agreement with stochastic
simulations carried out on a square lattice with periodic
boundaries. 

In the class of models studied here
an equilibrium configuration
with both types of cells  can only occur 
when $\kappa=1$.
An oft quoted result for the WB model is that the regression probability
is $\frac{1}{\kappa}$ (or $\frac{1}{\kappa+1}$).
 Our simulations and analysis show that
the result $\frac{1}{\kappa+1}$ applies both to interface implementations
of the WB model and to the full WB model.

In subsequent work we plan to explore the important question of the rate
of approach to the equilibrium state.

This work has been supported by the Australian Research Council.

\newpage

\newpage

\begin{thebibliography}{9}
\bibitem{FN1} Under appropriate conditions the
bacterial front may also spread via a Laplacian growth process.
\bibitem{FN2} It would be straightforward to modify the model to
allow for the possibility that when a cell in the basal layer
divides the daughter cell may be pushed directly up out of the basal layer.
\end{thebibliography}
\end{document}